\documentclass[a4paper]{article}

\usepackage{18lomcon}        
\usepackage{cite}             
\usepackage{epsfig}           
\usepackage{epstopdf}

\def\rate{counts/(keV$\cdot$kg$\cdot$y)~}

\begin{document}


\title{THE CUORE EXPERIMENT AT LNGS }

\author{Davide Chiesa \email{davide.chiesa@mib.infn.it} on behalf of the CUORE collaboration}

\affiliation{Dipartimento di Fisica, Universit\`a di Milano-Bicocca, Milano I-20126 - Italy \\
INFN - Sezione di Milano Bicocca, Milano I-20126 - Italy}

\date{}
\maketitle


\begin{abstract}
The Cryogenic Underground Observatory for Rare Events (CUORE) is the first bolometric experiment searching for neutrinoless double beta decay that has been able to reach the 1-ton scale. The detector consists of an array of 988 TeO$_2$ crystals arranged in a cylindrical compact structure of 19 towers. The construction of the experiment and, in particular, the installation of all towers in the cryostat was completed in August 2016, followed by the cooldown to base temperature at the beginning of 2017. The CUORE detector is now operational and has been taking science data since Spring 2017. We present here the initial performance of the detector and the preliminary results from the first detector run.
\end{abstract}

\section*{}

CUORE (Cryogenic Underground Observatory for Rare Events) is an experiment that searches the neutrinoless double beta decay ($0\nu\beta\beta$) of $^{130}$Te~\cite{CUOREexp}. This nuclear transition, if observed, would imply that lepton number is not conserved and that neutrinos are Majorana particles~\cite{Majorana}. 

\begin{figure}[htb!]
  \begin{minipage}{5.9cm}
     \centering
     \includegraphics[width=5cm]{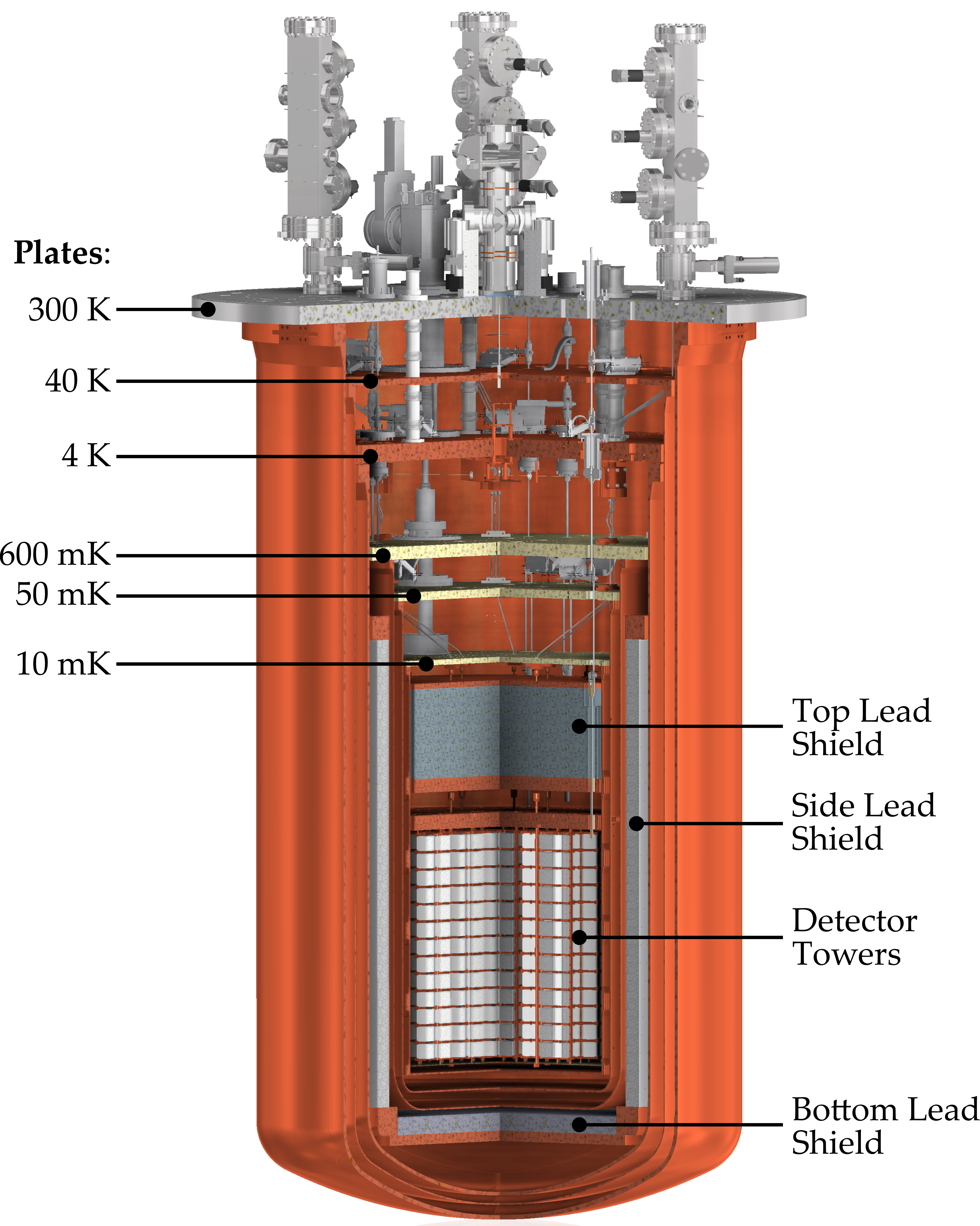}
  \end{minipage}
  \begin{minipage}{5.9cm}
     \centering
     \includegraphics[width=5cm]{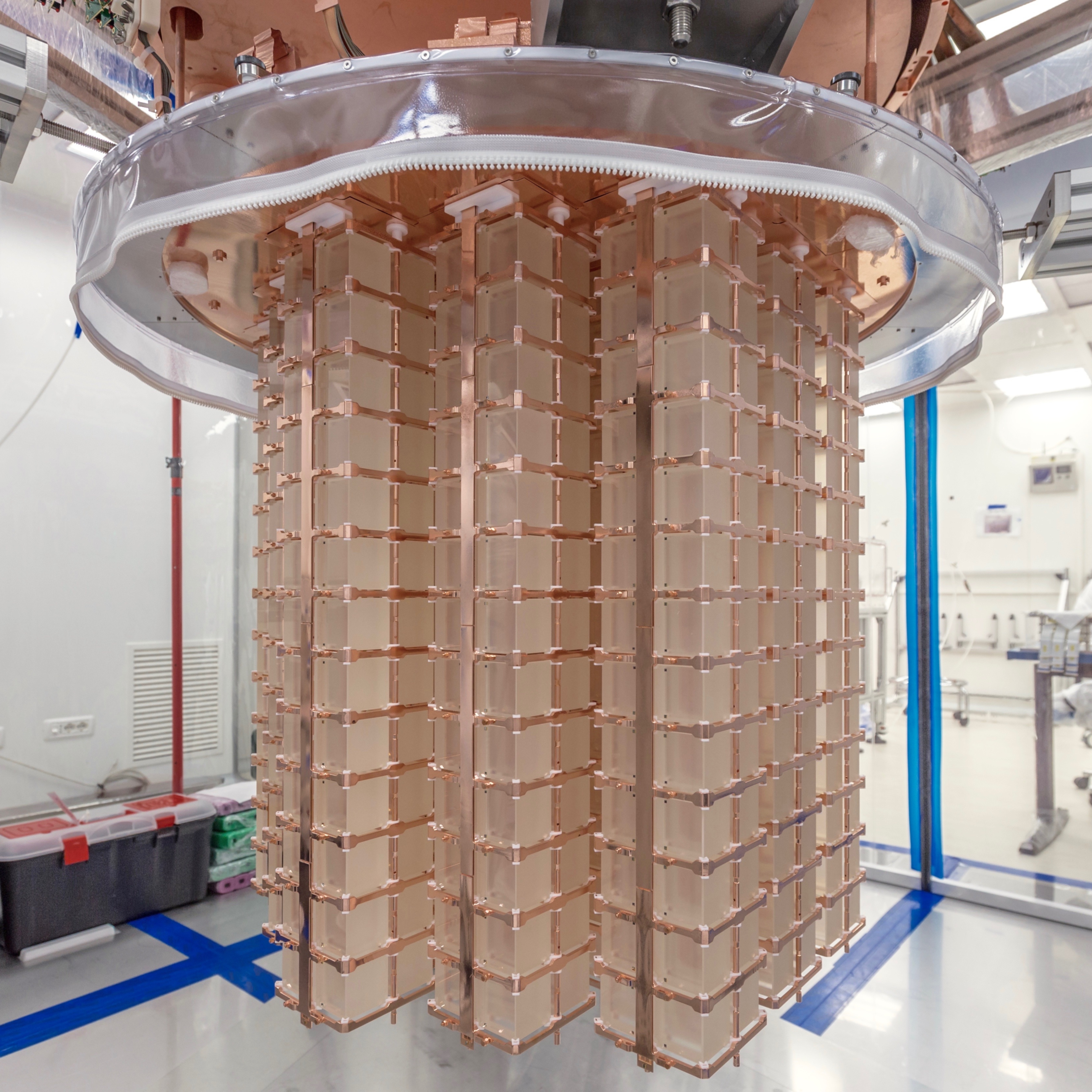}
  \end{minipage}
\caption{An illustration of the CUORE cryostat (left) and a picture of the CUORE detector (right).}
\label{Fig:Detector}
\end{figure}

The CUORE detector (Fig.~\ref{Fig:Detector}) is composed by 988 TeO$_2$ bolometers operated at a temperature of $\sim$10~mK in a cryostat installed at \textit{Laboratori Nazionali del Gran Sasso} (LNGS) in Italy. Each bolometer is a cubic TeO$_2$ crystal with 5~cm side and 750~g average mass, equipped with a neutron transmutation doped (NTD) thermistor and a silicon heather. The TeO$_2$ crystals are made with natural tellurium, that includes $^{130}$Te with 34.2\% isotopic abundance. The 988 bolometers are arranged in a closely packed array of 19 towers, each consisting of 13 floors of 4 crystals. With a total detector mass around 740~kg of TeO$_2$ (206~kg of $^{130}$Te), CUORE is the first ton-scale cryogenic detector for the search of $0\nu\beta\beta$ decay.

The CUORE detector measures the energy released inside the bolometers by particle interaction. The experimental signature of the $0\nu\beta\beta$ decay is a peak of events at the $Q$-value of the transition ($Q_{\beta\beta}$=2528~keV for $^{130}$Te). Since the $0\nu\beta\beta$ decay is an extremely rare decay, the challenge of CUORE is to maximize its experimental sensitivity. For this purpose, the CUORE detector is designed to reach good energy resolution and low background rate in the $0\nu\beta\beta$ Region of Interest (ROI). With an energy resolution of 5~keV at $Q_{\beta\beta}$ and a background rate of $10^{-2}$\rate in the ROI, CUORE is expected to reach an experimental sensitivity of $T_{1/2}^{0\nu}>9\times10^{25}$yr (90\% C.L.) with 5~yr of live time~\cite{Sensitivity}.

The materials used to build the CUORE detector and its cryostat were selected with the aim of minimizing the radioactive contaminations that can produce background events in the ROI. Moreover, the detector components were produced, handled and cleaned according to specific protocols, and the detector installation was performed in a radon free environment to avoid any possible recontamination. The background due to muons is strongly suppressed thanks to the experiment location at LNGS, with $\sim$1400 of overlying rock. The environmental background due to external $\gamma$-rays and neutrons is shielded by different layers of copper, lead, borated polyethylene and boric acid, that are included in the cryostat itself or surround it.
To evaluate the expected background in the ROI, we developed a detailed Monte Carlo simulation that exploits the information about the contaminations of materials obtained through radio-assay screening campaigns and bolometric measurements~\cite{BkgBudget, 2nbb}, proving that the goal of a background rate of $10^{-2}$\rate is within the reach.

The installation of the 19 towers was successfully completed in Summer 2016, obtaining 984 functioning bolometers out of 988. The cryostat interfaces and radiation shields were assembled in the following months. At the beginning of 2017, we started the detector pre-operation phase, to optimize the signal readout and the working points of the bolometers. 
In May 2017, we collected three weeks of physics data bracketed by two calibration periods. 
The detector stability has been much improved compared to Cuoricino and CUORE-0 predecessor experiments~\cite{Qino,Q0} and, for the first analysis, we acquired an exposure of 38.1~kg$\cdot$yr of TeO$_2$ (10.6~kg$\cdot$yr of $^{130}$Te). 
During calibration measurements, 12 strings populated with $^{232}$Th sources were temporarily deployed inside the detector region. Six $\gamma$-lines of $^{232}$Th decay chain (from 239~keV to 2615~keV) are then used to perform the energy calibration of the detected pulses.

\begin{figure}[t!]
  \begin{minipage}{6.7cm}
     \centering
     \includegraphics[width=6.6cm]{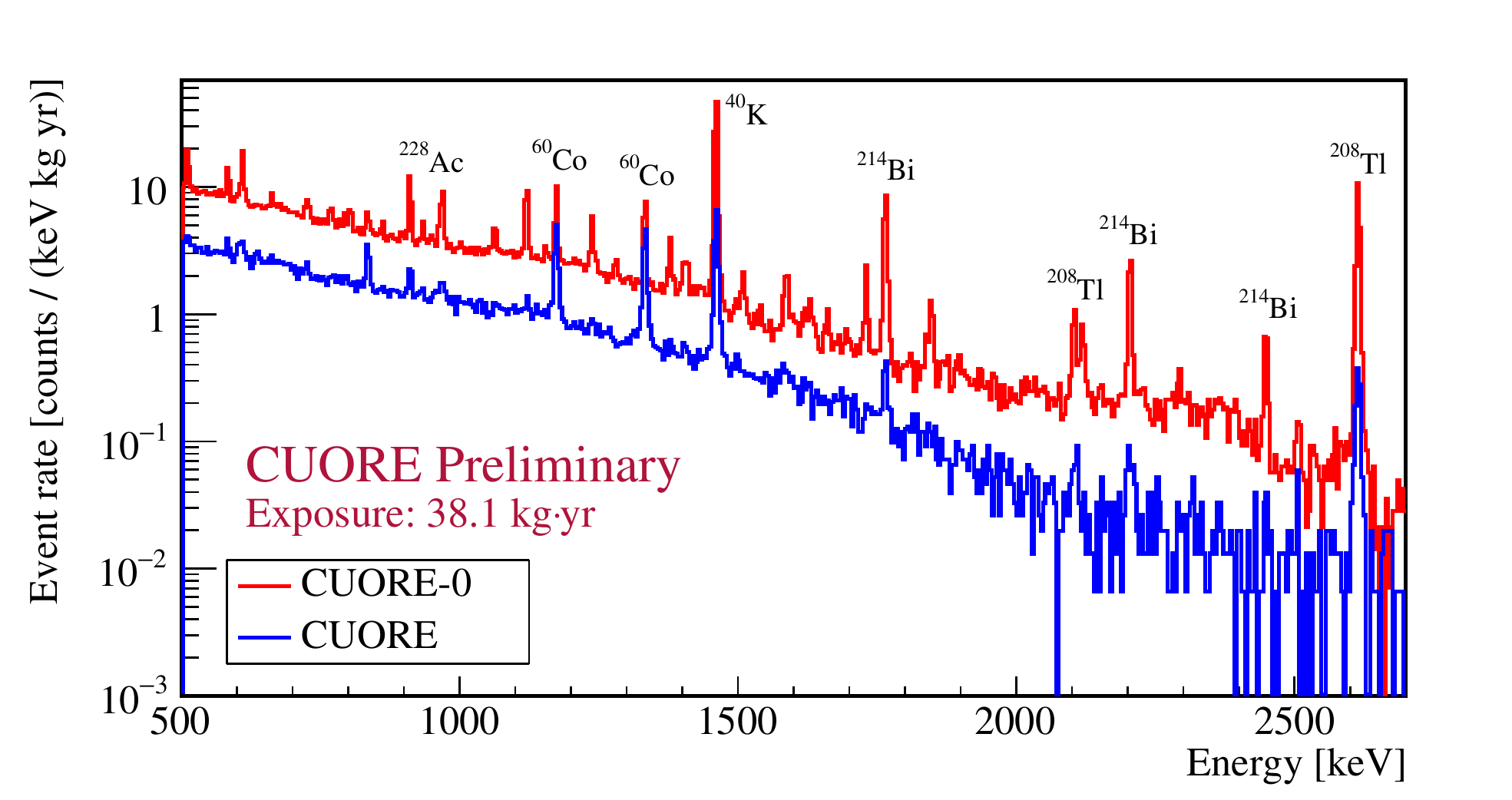}
  \end{minipage}
  \begin{minipage}{5.2cm}
     \centering
     \includegraphics[width=5.2cm]{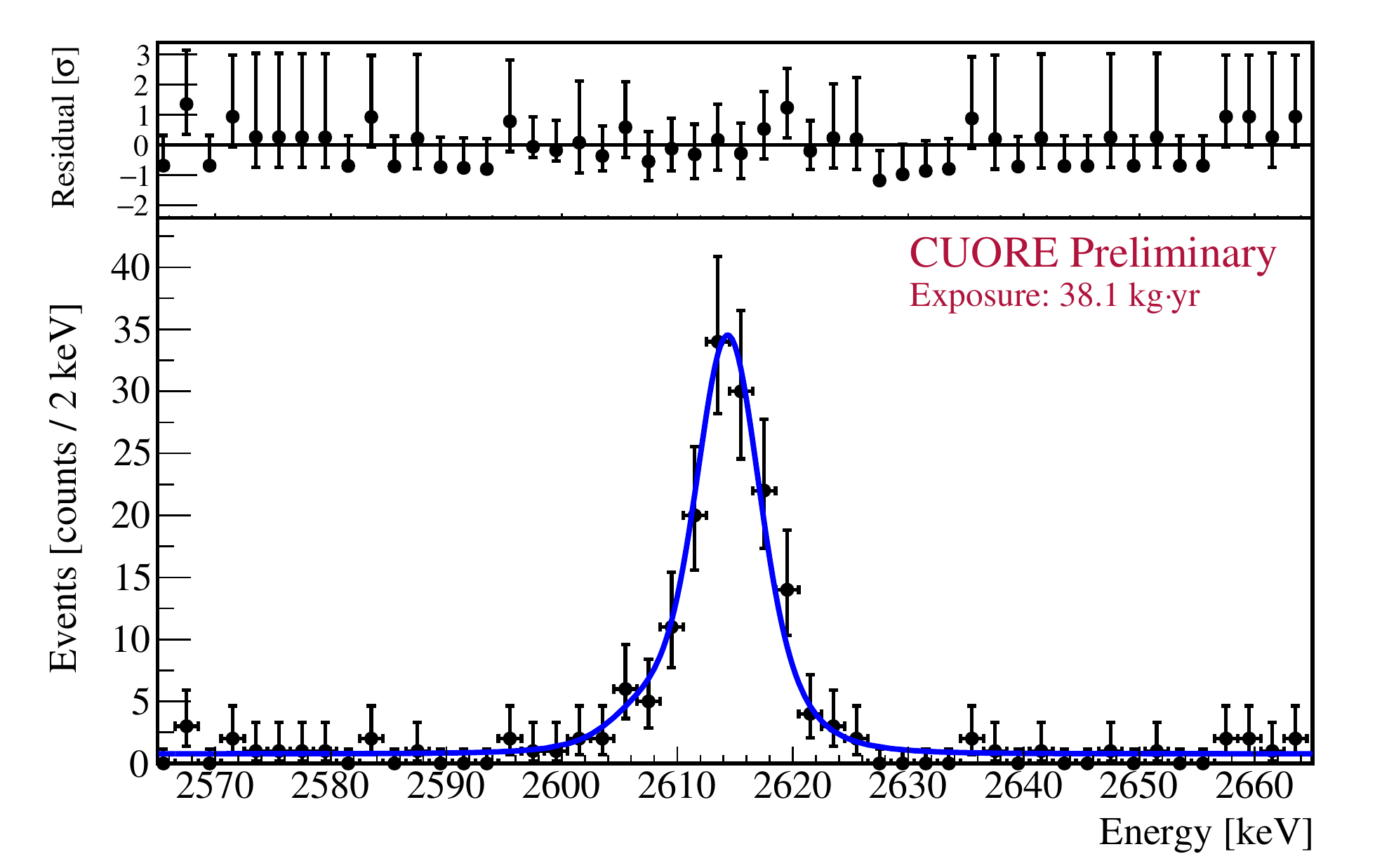}
  \end{minipage}
\caption{Left: comparison of physics spectra in the gamma region measured with CUORE and CUORE-0, with prominent $\gamma$-lines labelled. Right: detail of the 2615~keV line.}
\label{Fig:BkgFWHM}
\end{figure}

The physics spectrum (Fig.~\ref{Fig:BkgFWHM}, left) is built after applying a series of selection criteria aimed at improving the experimental sensitivity. First, we remove periods of low quality data --caused by noisy laboratory conditions-- and we reject pile-up events. Then, we select only signals consistent with a proper template waveform (pulse shape analysis) in order to identify real particle events. Finally, we exclude events that simultaneously trigger more than one crystal, to reduce the background due to events depositing energy in multiple crystals. We evaluate an overall detection efficiency of (55.3$\pm$3.0)\%, which includes the (88.35$\pm$0.09)\% probability that a $0\nu\beta\beta$ decay is fully contained in a single crystal and the (62.6$\pm$3.4)\% probability that a physics event is not discarded when the selection criteria are applied.

We evaluate the detector energy resolution near the ROI by fitting the 2615~keV line in the physics spectrum. The armonic mean of the detector FWHM resolutions is 7.9$\pm$0.6~keV (Fig.~\ref{Fig:BkgFWHM}, right).

\begin{figure}[htb!]
  \begin{minipage}{5.9cm}
     \centering
     \includegraphics[width=5.9cm]{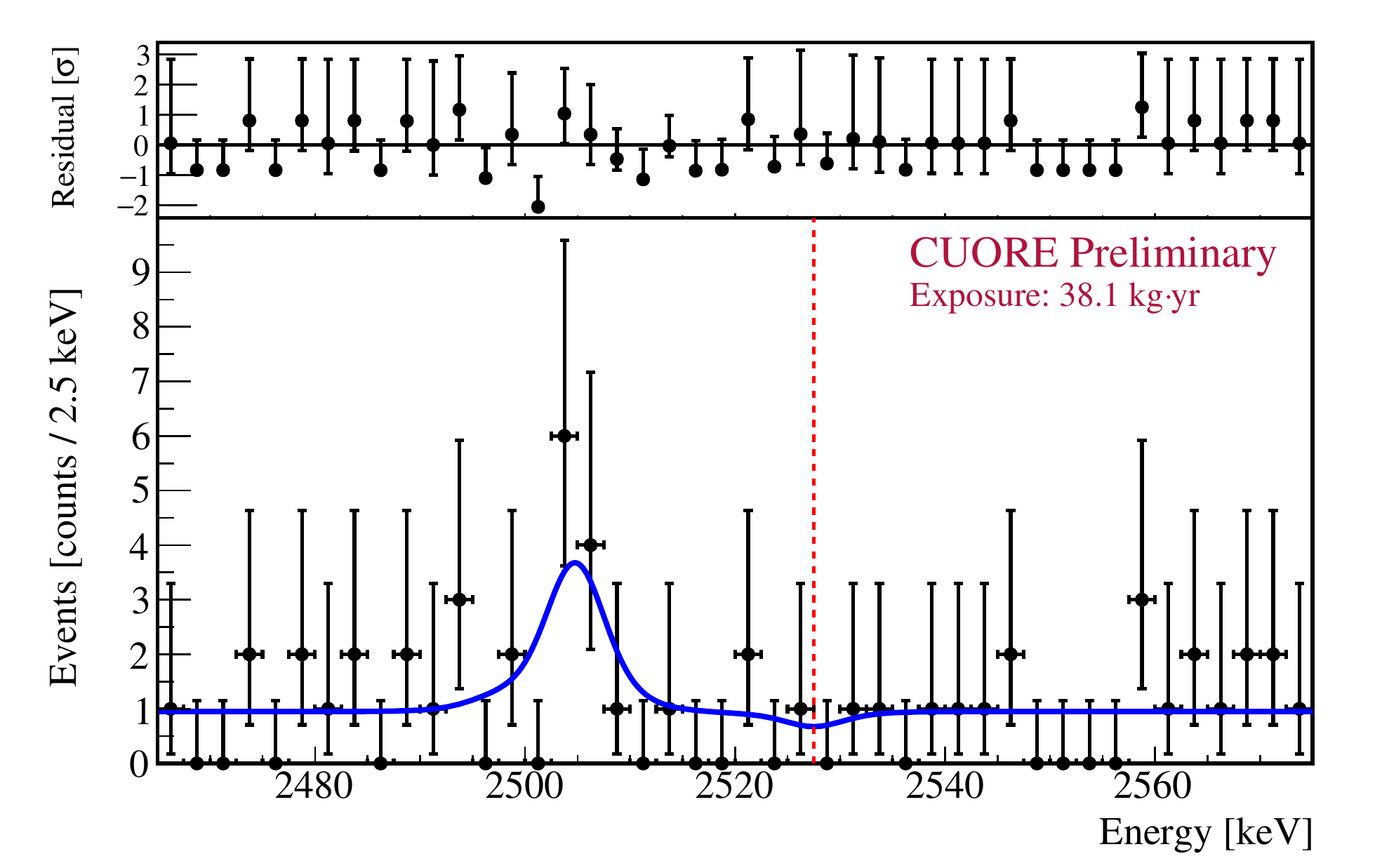}
  \end{minipage}
  \begin{minipage}{5.9cm}
     \centering
     \includegraphics[width=5.9cm]{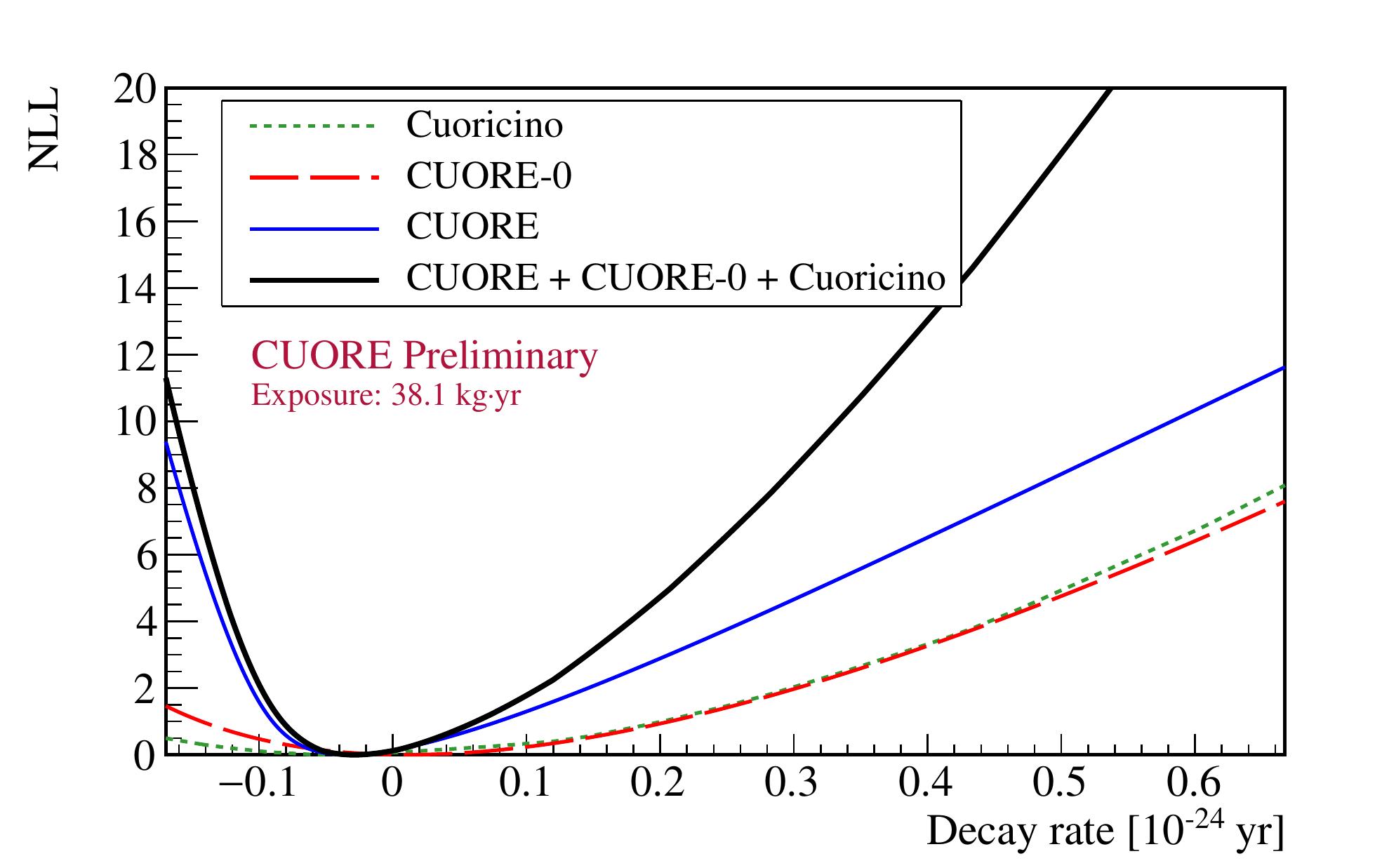}
  \end{minipage}
\caption{Left: Fit of the CUORE spectrum in the ROI. The peak near 2506~keV is attributed to $^{60}$Co. Right: Profile negative-log-likelihood curves for CUORE, CUORE-0, Cuoricino, and their combination.}
\label{Fig:Fit}
\end{figure}

To estimate the background in the ROI and the $0\nu\beta\beta$ decay rate ($\Gamma_{0\nu}$), we perform an Unbinned Extended Maximum Likelihood fit in the [2465--2575]~keV range (Fig.\ref{Fig:Fit}, left) with the same procedure used for CUORE-0~\cite{Q0Analysis}.
The best-fit values are $0.98^{+0.17}_{-0.15}\times10^{-2}$\rate for the background rate, and $(-0.03^{+0.07}_{-0.04}$(stat.)$\pm0.01$(syst.))$\times10^{-24}$yr$^{-1}$ for $\Gamma_{0\nu}$.
We find no evidence for the $0\nu\beta\beta$ decay of $^{130}$Te and we can only calculate an upper limit of $\Gamma_{0\nu}$, by integrating the profile likelihood in the physical region ($\Gamma_{0\nu}\geq0$). This corresponds to a half-life lower limit of $T_{1/2}^{0\nu}>4.5\times10^{24}$yr (90\% C.L.).

Finally, we combine the first results from CUORE with those obtained from CUORE-0~\cite{Q0} and Cuoricino~\cite{Qino} with 9.8~kg$\cdot$yr and 19.8~kg$\cdot$yr exposure of $^{130}$Te, respectively (Fig.~\ref{Fig:Fit}, right). The half-life lower limit obtained by combining the profile negative-log-likelihood curves of the three experiments is $T_{1/2}^{0\nu}>6.1\times10^{24}$yr (90\% C.L.).

In summary, CUORE is the first ton-scale cryogenic detector array in operation and with three week of physics data we were able to set the most stringent limit on the $0\nu\beta\beta$ half-life of $^{130}$Te, surpassing the previous one obtained by combining the results from CUORE-0 and Cuoricino experiments. From the analysis of the first dataset, we got important information about detector performances, that allow to confirm the expected sensitivity of CUORE for the search of the $0\nu\beta\beta$ decay of $^{130}$Te.

\noindent\textit{Note}: A new science run was carried out during August 2017. The new dataset and the one described in this paper were re-processed with a slightly improved analysis procedure. The results were submitted for publication to PRL~\cite{CUORE_ArXiV}.

\section*{Acknowledgments}

The CUORE Collaboration thanks the directors and staff of the Laboratori Nazionali del Gran Sasso and the technical staff of our laboratories. This work was supported by the Istituto Nazionale di Fisica Nucleare (INFN); the National Science Foundation; the Alfred P. Sloan Foundation; the University of Wisconsin Foundation; and Yale University. This material is also based upon work supported by the US Department of Energy (DOE) Office of Science; and by the DOE Office of Science, Office of Nuclear Physics.  This research used resources of the National Energy Research Scientific Computing Center (NERSC).


\end{document}